# Probing excitonic states in ultraclean suspended two-dimensional semiconductors by photocurrent spectroscopy

A. R. Klots[1,†], A. K. M. Newaz[1, †], Bin Wang[1], D. Prasai[2], H. Krzyzanowska[1], D. Caudel[1], N. J. Ghimire[3,4], J. Yan[4,5], B. L. Ivanov[1], K. A. Velizhanin[6], A. Burger[7], D. G. Mandrus[3,4,5], N. H. Tolk[1], S. T. Pantelides[1,4], and K. I. Bolotin[1,*]

[1]Department of Physics and Astronomy, Vanderbilt University, Nashville, TN-37235, USA
[2]Interdisciplinary Graduate Program in Materials Science, Vanderbilt University, Nashville, TN-37234, USA
[3]Department of Physics and Astronomy, University of Tennessee, Knoxville, TN-37996, USA
[4]Materials Science and Technology Division, Oak Ridge National Laboratory, Oak Ridge, TN-37831, USA
[5]Department of Materials Science and Engineering, University of Tennessee, Knoxville, TN-37996, USA
[6]Theoretical Division, Los Alamos National Laboratory, Los Alamos, NM-87545, USA
[7]Department of Physics, Fisk University, Nashville, TN-37208, USA

([†]These authors contributed equally to the work, [*]email: kirill.bolotin@vanderbilt.edu)

Date: March 26, 2014

The optical response of semiconducting monolayer transition-metal dichalcogenides (TMDCs) is dominated by strongly bound excitons that are stable even at room temperature. However, substrate-related effects such as screening and disorder in currently available specimens mask many anticipated physical phenomena and limit device applications of TMDCs. Here, we demonstrate that that these undesirable effects are strongly suppressed in suspended devices. Extremely robust (photogain >1,000) and fast (response time <1ms) photoresponse combined with the high quality of our devices allow us to study, for the first time, the formation, binding energies, and dissociation mechanisms of excitons in TMDCs through photocurrent spectroscopy. By analyzing the spectral positions of peaks in the photocurrent and by comparing them with first-principles calculations, we obtain binding energies, band gaps and spin-orbit splitting in monolayer TMDCs. For monolayer $MoS_2$, in particular, we estimate an extremely large binding energy for band-edge excitons, $E_{bind} \geq 570$meV. Along with band-edge excitons, we observe excitons associated with a van Hove singularity of rather unique nature. The analysis of the source-drain voltage dependence of photocurrent spectra reveals exciton dissociation and photoconversion mechanisms in TMDCs.

Monolayer (1L) transition metal dichalcogenides (TMDCs), such as molybdenum disulfide ($MoS_2$), molybdenum diselenide ($MoSe_2$), or tungsten diselenide ($WSe_2$) are two-dimensional atomic crystals. In contrast to graphene[1], a prototypical 2D material, 1L-TMDCs are direct band gap semiconductors with strong spin-orbit interactions, which cause spin-splitting of the valence band of TMDCs[2-4] and allow optical manipulation of spin- and valley- degrees of freedom in these materials[5-8]. Two-dimensional confinement, high effective carrier mass and weak screening lead to strong electron-electron interactions and dominance of tightly bound excitons in the optical properties of 1L-TMDCs[2-11]. These extraordinary properties make TMDCs ideal platform for studying many anticipated phenomena including quantum-, valley- and spin-Hall effects[3,12], superconductivity in monolayer $MoS_2$[13,14] and many-body effects[9,10,15]. Moreover, strong light-matter interactions[16] make TMDCs excellent materials for ultrasensitive photodetectors[17] and energy harvesting devices[18]. Despite rapid progress in understanding the electronic and optical properties of TMDCs[19], important fundamental questions remain unanswered:

1) How do substrate-related effects perturb the intrinsic properties of monolayer TMDCs? Indeed, there are indications that the presence of a substrate can cause strong carrier scattering[20,21] and affect exciton energies through screening[22].

2) What types of excitons exist in TMDCs and what are their binding energies? While calculations predict a plethora of excitonic states with extremely large binding energies[23,24], experimental progress has been hampered by large broadening of the excitonic peaks in the available samples[2,23].

3) What are the photoconversion mechanisms in TMDC devices? Despite indications of efficient photoconversion[16,17], photodetection[17], and strong interest in employing TMDCs as solar cells[18], it is currently unclear how strongly-bound excitons in TMDCs dissociate and contribute to the photocurrent.

Here, we present our experimental results answering these questions. First, we significantly reduce disorder and eliminate substrate-related screening in TMDCs by fabricating free-standing and electrically contacted $MoS_2$, $MoSe_2$, and $WSe_2$ specimens. We then use photocurrent spectroscopy as a versatile tool for studying excitons and their dissociation mechanisms. In monolayer (1L) $MoS_2$, we have observed well-defined peaks at ~1.9 eV and ~2.1 eV ('A' and 'B') and a broad peak 'C' at ~2.9 eV. We attribute the peaks A and B to optical absorption by band-edge excitons, and the peak C to absorption by excitons associated with the van Hove singularity of $MoS_2$. Compared to previously reported optical absorption measurements of supported $MoS_2$[2], our photocurrent spectra exhibit sharp and isolated peaks with near-zero background between them, suggesting the absence of disorder-related midgap states. High quality of our devices allows to obtain experimentally, for the first time, the lower bound of the binding energy of band-edge excitons of $MoS_2$, $E_{bind} \geq 570$ meV. Finally, we investigate the photoconversion and photogain mechanisms in monolayer TMDCs. By controlling the source-drain voltage, we observe different dissociation pathways for A/B- and C-excitonic states, demonstrate photogain of the order of 1000 with response times faster than 1 ms, and uncover the mechanism of this photogain. We also demonstrate the universality of our techniques by performing measurements on other materials, such as bi- and multi-layer $MoS_2$, monolayer $MoSe_2$ and monolayer $WSe_2$. Our results demonstrate, for the first time, that photocurrent spectroscopy is an efficient tool for probing single- and many-body states in pristine TMDCs and suggest the application of TMDCs as efficient photodetectors with a voltage-tunable spectral response.

To decrease the substrate-induced disorder in TMDCs, we fabricate electrically contacted suspended devices (see Supplementary information, S1 for details), following the approach developed for graphene[20]. Initially, we focus on 1L-MoS$_2$ devices (Fig. 1a, Inset), while discussing the case of monolayer MoSe$_2$, WSe$_2$, and multilayer MoS$_2$ later. Two-probe electrical transport measurements indicate that upon suspension the field effect carrier mobility ($\mu$) of a typical device, ~0.05 cm$^2$/Vs, increases by an order of magnitude (Fig. 1a), consistent with a recent report[21]. We note that since neither the contact resistance nor the carrier density can be determined in the two-probe geometry, the physically relevant Hall mobility of the same device may be larger by orders of magnitude[25,26]. To further increase the quality of suspended devices, we rely on thermal annealing, which is effective in improving $\mu$ both for graphene[20] and multilayer MoS$_2$[27]. Since the low electrical conductance ($G$) of MoS$_2$ devices precludes annealing via Ohmic heating[20], we instead locally heat the region of the wafer that is in thermal contact with the device. The annealing is performed *in situ* inside a cryostat kept at base temperature $T=77K$ using a ~5W CO$_2$ laser beam, which is defocused (intensity <20μW/μm$^2$) to avoid sample damage. This annealing renders the device near-insulating under small source-drain bias voltage $|V_{ds}|<1V$ (Fig. 1a, red curve, and Fig. 1b, black curve). This behavior is expected for a pristine undoped semiconductor with the Fermi level located inside the band gap. Since the gate voltage is limited to $|V_g|<12V$ to avoid electrostatic collapse of MoS$_2$, we are unable to achieve either electron or hole conductivity regimes via electrostatic gating.

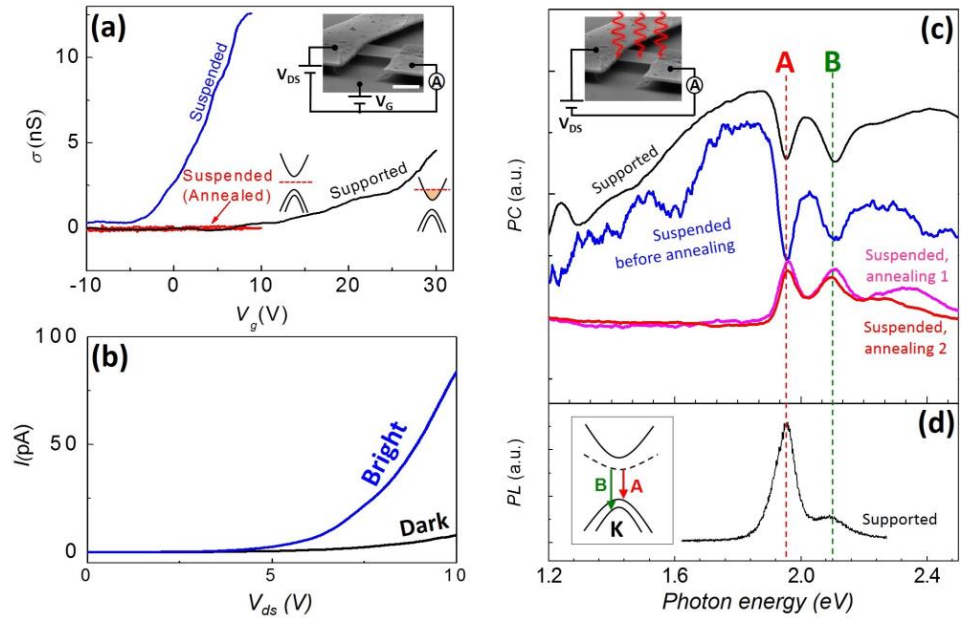

**Figure 1: Effects of thermal annealing on conductance and photocurrent of suspended MoS$_2$.** **(a)** Gate-dependent conductance of supported, suspended, and suspended annealed 1L-MoS$_2$ devices at $T=300K$. Inset: Image of the device. The scale bar is 1 μm. Schematically drawn band diagrams show the position of the Fermi level (red dashed line). **(b)** Dark and bright electrical response of an annealed suspended device at $T=77K$. Illumination power is $P$~3pW/μm$^2$ and wavelength is $\lambda=430$nm. **(c)** Photocurrent (PC) spectrum of a supported and suspended MoS$_2$ devices at different stages of thermal annealing at $T=77K$. **(d)** Photoluminescence spectra for a supported MoS$_2$ device at $T=300K$. Since PL spectra were recorded at room temperature, we manually blue-shift them by 150meV to allow comparison with PC spectra obtained at $T=77K$ (see Supplementary Information, S4 for details). Inset: Bandstructure schematics of MoS$_2$ near $K$-point illustrating the origin of band-edge excitons. The dashed line represents excitonic states.

To investigate suspended and annealed devices further, we measure PC under high $V_{ds}$ (>3V) (Fig. 1b, blue curve). We illuminate the entire device using a low intensity ($P \leq 30$ pW/μm²) light source and record photocurrent $I_{PC}$ across the device as a function of the photon energy $\hbar\omega$ (Fig. 1c). The total current through the device is $I = V_{ds} G(V_{ds}, n)$, where $G$ in turn depends on the number of charge carriers $n$ and $V_{ds}$. Upon illumination with power $P$, $n$ increases by $\Delta n = (P/\hbar\omega)\alpha(\hbar\omega)D\tau$, where $\alpha$ is the absorption coefficient, $D$ is the photoconversion probability (the probability of generating an unbound photocarrier by an absorbed photon), and $\tau$ is the photocarrier lifetime[28]. For a constant $V_{ds}$, the photocurrent is

$$I_{PC} = V_{ds}\frac{\partial G}{\partial n}\Delta n = \left[V_{ds}\frac{\partial G}{\partial n}\frac{D\tau}{e}\right]e\frac{P}{\hbar\omega}\alpha(\hbar\omega) \qquad (1),$$

where $e$ is the electron charge. The expression inside the brackets is the photogain $\eta$, the ratio between the number of absorbed photons and the number of photocarriers transported across the device per unit time. For the device presented in Fig. 1b, we estimate $\eta \sim 200$ at $V_{ds} \sim 10$V, and in other devices (Fig. 4b) $\eta > 1,000$ (we assume that $\alpha(1.9\text{eV}) \sim 0.1$ and $\alpha(2.9\text{eV}) \sim 0.4^2$).

Equation (1) is central to the analysis of our data as it shows that PC can be used to estimate the intrinsic parameters of TMDCs – $\alpha(\hbar\omega)$, $\tau$, and $D$. Indeed, since the photogain is weakly wavelength-dependent, peaks in $I_{PC}$ are associated with peaks in $\alpha(\hbar\omega)$ (See the Supplementary Information, S3 for more detail). On the other hand, the amplitude of $I_{PC}$ is related to photogain, and hence to $D$ and $\tau$. Therefore, similarly to optical absorption measurements, PC spectroscopy allows us to study single- and many-body electronic states in TMDCs[29,30]. Unlike absorption spectroscopy, PC can be easily measured for an electrically contacted microscopic device in a cryogenic environment, as the device itself acts as its own photodetector. Moreover, high photosensitivity of TMDC phototransistors allows us to use very low illumination intensity in our experiments, thereby excluding artifacts, such as photo-thermoelectric effects[31] (which would yield currents <0.1pA, more than three order of magnitude smaller than the photocurrent measured in our devices) and optically non-linear[32] effects, arising at high photocarrier densities. We first use PC spectroscopy to probe absorption spectrum $\alpha(\hbar\omega)$ of TMDCs, while later investigating the origins of large photogain.

For substrate-supported and suspended unannealed devices, we observe two dips (that were reported in bulk TMDCs previously[33]) at ~1.9eV and ~2.1eV on top of a largely featureless device-dependent background photocurrent (Fig. 1c) (detailed discussion is in the Supplementary Information, S2). Upon annealing, this background, attributable to absorption by midgap states[34] as well as to photogating artifacts[35,36], recedes leaving a set of universal features seen in every device (Fig. 2a). We observe: (*i*) Two sharp peaks at ~1.9 eV and ~2.1 eV (labeled 'A' and 'B', respectively), (*ii*) near-zero PC signal below the A-peak, between A- and B-peaks and above the B-peak (from ~2.1 eV to ~2.5eV), (*iii*) steep growth of PC starting at ~2.5eV, and (*iv*) a broad and strong peak 'C' at ~2.9eV. To the best of our knowledge, this is the first observation of the features *(ii)-(iv)* in PC spectroscopy. Next, we demonstrate that all of these features originate from optical absorption by bound excitons as well as by unbound electron-hole (*e-h*) pairs in $MoS_2$.

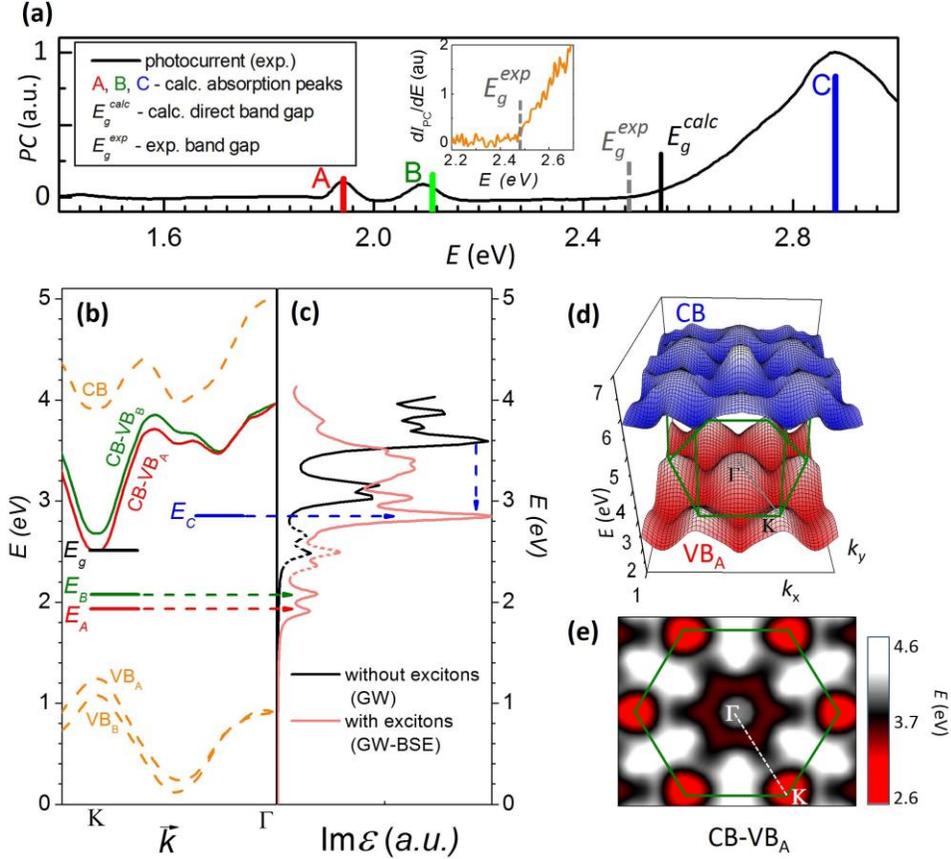

**Figure 2: Probing excitons in pristine monolayer MoS$_2$ through photocurrent spectroscopy.** **(a)** PC spectrum of an intrinsic suspended 1L-MoS$_2$ device. Calculated positions of excitonic A-, B- and C-peaks and band gap $E_g$ are shown as colored vertical bars. The bar height represents peaks amplitudes. The inset: derivative of the photocurrent plotted *vs.* the photon energy. **(b)** Electronic and optical band structures of 1L-MoS$_2$ along the *K-Γ* direction. The solid horizontal lines are the estimated positions of the excitonic bound states. **(c)** Optical spectrum of MoS$_2$ calculated with and without excitonic effects. The dashed peaks between 2.2 eV and 2.7 eV are computational artifacts, which are discussed in the Supplementary Information, S5. Vertical blue arrow indicates the position of the van Hove singularity downshifted by excitonic effects. **(d)** Three-dimensional plot of the band structure of MoS$_2$. **(e)** The colorplot of the optical band structure of MoS$_2$. Dark red gear shaped region around *Γ* is the local minimum corresponding to the excitonic C-peak.

Features A and B stem from optical absorption by the well-known[2,4,27] A- and B- band edge excitons of MoS$_2$ residing at *K*-points of the Brillouin zone (Fig. 1d, Inset). Recombination of these excitons results in photoluminescence peaks at similar spectral positions (Fig. 1d). The ~160 meV separation between the A- and B- peaks is a consequence of the splitting of the valence band of MoS$_2$ at the *K* point due to spin-orbit interactions[2-4]. The positions of the A- and B-peaks are also in good agreement with the calculated optical spectrum that we obtain using first-principles GW-BSE calculations (Fig. 2c, light-red curve,)[24,37-39]. See Supplementary Information, S5 for details.

The feature at ~2.9eV ('C') has been previously noted in absorption spectrum of MoS$_2$[2,27,37], but to the best of our knowledge not thoroughly analyzed. We interpret this peak as coming from an excitonic state associated with the van Hove singularity of 1L-MoS$_2$. This van Hove singularity is peculiar, as neither the conduction nor the valence bands have singularities in the density of states in the corresponding region of the Brillouin zone between $K$ and $\Gamma$ points (orange curves in Fig. 2b and Fig. 2d). At the same time, the bands are locally parallel in that region, causing a local minimum in the Mexican-hat-like *optical band structure* (difference between conduction and valence bands shown in Fig. 2b as red and green curves). This minimum is prominent in a 2D colorplot of the optical band structure as a continuous gear-shaped region circling the $\Gamma$ point (Fig. 2e, dark red region). The large joint density of states associated with this minimum yields a strong peak in $\alpha(\hbar\omega)$. Indeed, our GW calculations (*i.e.*, without inclusion of excitonic effects) of the optical spectrum prominently feature a sharp peak at ~3.45 eV, the value that corresponds to the optical band gap at the van Hove singularity point (Fig. 2c, black curve). Excitonic effects downshift the peak to ~2.9eV (Fig. 2c, light-red curve), very close to the experimentally measured position of the C-peak. Interestingly, the C-exciton valley of the optical bandstructure is near-rotationally symmetric rendering this exciton effectively one-dimensional[40]. Moreover, the location of the C-exciton at the bottom of the Mexican hat dispersion suggest that this exciton is localized in both real and momentum space, a conclusion also supported by first-principles calculations[23,37]..

Within the resolution of our measurements (signal-to-noise ratio is ~20 for A/B-peaks), we observe zero photocurrent below the A-peak, between the A- and B-peaks and between the B- and C-peaks. This observation is in contrast with non-zero optical absorption[2] and photocurrent in the same region in supported devices measured by us (data in the Supplementary Information, S3) as well as by others[2,27]. It has been previously suggested[41] and observed[27,42] that disorder-related midgap states can significantly perturb the optical response of $MoS_2$ leading to below-band gap absorption. Moreover, reduction in the background absorption upon annealing, which is likely associated with reduced disorder, has been recently observed in chemically exfoliated $MoS_2$ samples[27]. We therefore interpret the lack of PC background in our devices as a signature of either the low density of the disorder-related midgap states or suppression of substrate-related screening. Moreover, we do not observe any features due to trions[9,10] and trapped excitons[42], which suggests that our devices are undoped and contain low defect density. We also note that despite the high quality of our devices, no signatures of anticipated[23,43] excited states of A or B excitons are observed. This is consistent with the very low oscillator strength of these states expected from a simple 2D hydrogen model (see Supplementary Information, S6).

Above the near-zero photocurrent region, we observe a featureless and abrupt increase of the PC above $E_g^{exp}$ ~2.5 eV. This increase is clearly visible in the plot of $dI_{pc}/d(\hbar\omega)$ (Fig. 2a, Inset). The PC onset occurs very close in energy to the calculated *fundamental* (i.e. single-particle) band gap of 1L-$MoS_2$, $E_g^{calc}$ ~2.55 eV (Fig. 2b-c) and is therefore related to direct band-to-band absorption by unbound $e$-$h$ pairs. However, experimentally we cannot distinguish the onset of the band-to-band absorption from the tail of the C-peak. We therefore interpret that the measured value of $E_g$ is a lower bound for the fundamental band gap value. We can therefore estimate the exciton binding energy in $MoS_2$ as $E_{bind}=E_g-E_A \geq 570$ meV. We emphasize that in our suspended devices the measured values for $E_g$ and $E_{bind}$ are free from the influence of the substrate-related dielectric screening and hence can be directly compared to calculations (Fig. 2a-c).

We now turn to bi- and multi-layer $MoS_2$, as well as other 1L-TMDCs, such as $MoSe_2$ and $WSe_2$. Similar A-, B-, and C- features are seen in photocurrent spectra for all of these materials (Fig. 3a). For materials other than 1L-$MoS_2$, however, we do not observe the zero photocurrent between B- and C-peaks. This precludes experimental estimation of exciton binding energies in these materials. However, similarly to the case of $MoS_2$, by obtaining good agreement between positions of A-, B- and C-peaks in experimental data and first-principles calculations, we can infer $E_g$ and $E_{bind}$ (details are in Supplementary Information, S5). We note the following trends:

(i) The A- and B- peaks in $MoS_2$ do not depend significantly on its thickness (Fig. 3b, red points)[2]. This is a consequence of simultaneous and nearly equal reduction of $E_g$ (Fig. 3b, black points) and $E_{bind}$ with the number of layers of $MoS_2$[44].

(ii) The splitting between A and B peaks is largest in $WSe_2$ (~510 meV), followed by $MoSe_2$ and $MoS_2$ (Fig. 3d). This is a signature of the stronger spin-orbit interaction in $WSe_2$, related to the higher atomic number of tungsten.

(iii) The calculations suggest that variation of the type of chalcogen (S, Se) atom has a strong effect on $E_g$ (Fig. 3c). This is a consequence of the dependence of the lattice constant on the type of chalcogen atoms. On the other hand, $E_{bind}$ remains roughly constant for all measured materials (Fig. 3d).

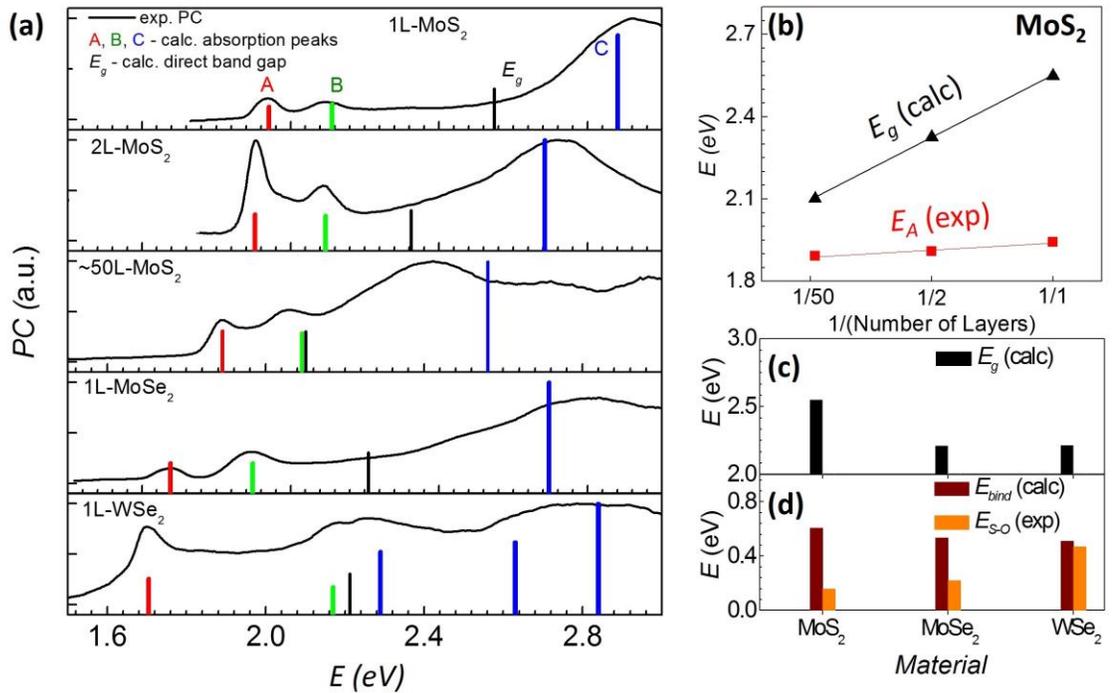

**Figure 3: Photocurrent in various TMDC materials.** (a) Experimental PC spectra of different TMDC devices. Solid bars are calculated excitonic peaks and band gap values. The $MoS_2$ device shown here is different from the device of Fig. 2. Large spin-orbit coupling of $WSe_2$ results in splitting of the valence and the conduction bands even near $\Gamma$-point, which leads to splitting of the C-peak. All the devices are suspended and annealed except for the multilayer $MoS_2$ device, which is supported on a glass substrate (see Supplementary Information, S1). (b) Dependence of excitonic peak positions and band gap values on number of layers of $MoS_2$. (c,d) Comparison of $E_g$, $E_{bind}$ and spin-orbit coupling strengths for different 1L-TMDCs.

Our next aim is to explain very large PC magnitude. To contribute to photocurrent, a neutral exciton must first dissociate into an unbound electron-hole pair. This process is characterized by the probability $D$ entering into Eq. (1). To investigate the mechanism of dissociation in 1L-$MoS_2$, we examine $I_{PC}$ vs. $V_{ds}$. We find that the A and B peaks in the photocurrent practically disappear at low $V_{ds}$, while the C peak remains prominent (Fig. 4a). This behavior is consistent with dissociation of excitons by strong electric fields arising near the interface between MoS$_2$ and metallic contacts. Indeed, a large electric field is required to overcome the binding energy $E_{bind} \geq 0.6$ eV for A-excitons. Such a field can arise at the interface between MoS$_2$ and a metallic contact due to the application of a large bias voltage (like in the case of pristine organic semiconductors[45]) and possibly due to the mismatch of the work functions of MoS$_2$ and metal (similar to nanotube devices[46] and excitonic solar cells[47]). Our conclusion that PC is produced only at the contacts is also supported by scanning photocurrent microscopy measurements directly mapping photocurrent production[48]. In contrast, C-excitons exist above the band gap and therefore can produce unbound e-h pair even without application of an external electric field. Thus our data are consistent with electric field assisted dissociation of A- and B-excitons and spontaneous decay of C-excitons into a free electron-hole pairs.

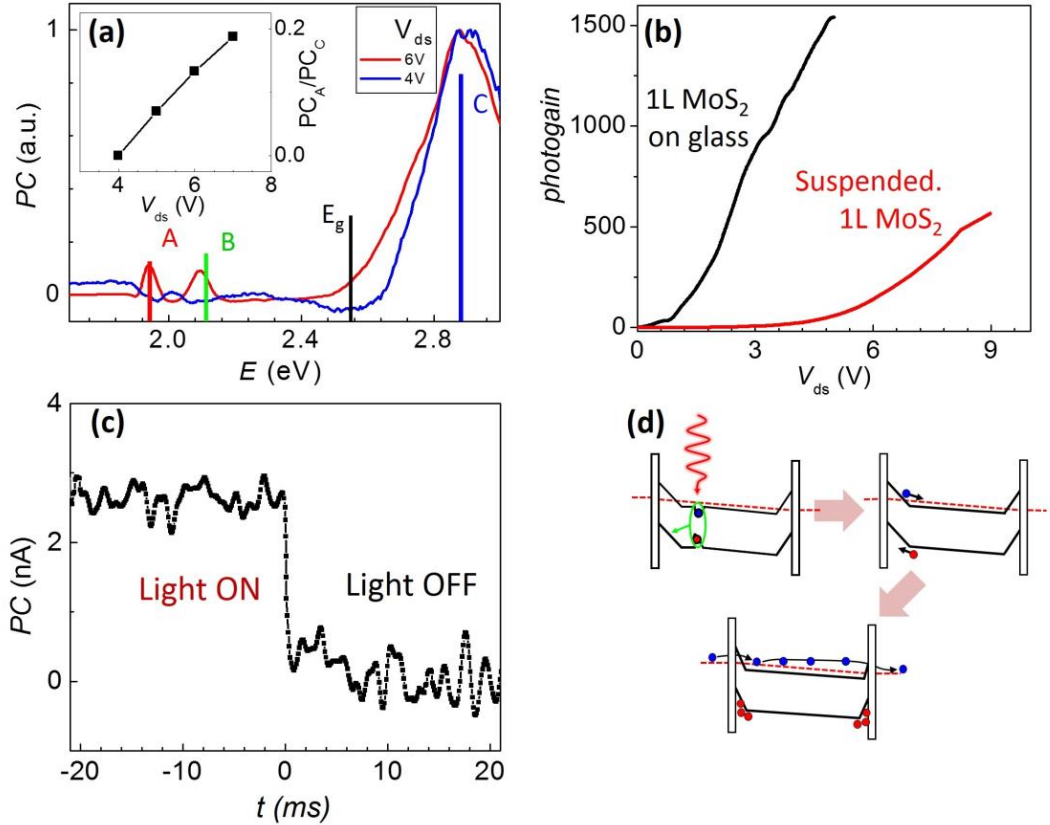

Figure 4: **Photoconversion mechanisms in monolayer MoS$_2$**. (a) PC spectra measured in a suspended 1L-MoS$_2$ at two different $V_{ds}$. Both curves are normalized to the height of the C-peak. Inset: relative PC amplitudes of A- and C- peaks *vs.* $V_{ds}$. (b) Photogain for a glass-supported and suspended devices *vs.* $V_{ds}$. The device is illuminated at λ=640nm with $P \sim 30$ pW/μm$^2$. (c) Time response of PC to the varying light intensity in a glass-supported MoS$_2$. This measurements sets the upper limit for the response time <1ms. Accuracy of time-resolved measurements was limited by the high resistance of MoS$_2$ and therefore high RC-time constant of the measurement circuit. (d) Schematics of the photogain mechanism.

Finally, we analyze the reason for the very large ($\eta > 1,000$) photogain in our devices (Fig. 4b). Previously suggested mechanisms, such as the direct dissociation at the contacts (yielding only $\eta < 1$)[48] or photothermoelectric effect (yielding $\eta << 0.1$)[31] cannot explain very high observed photogain. Generally, large gain can be related either to multiplication of photocarriers due to the avalanche effect[28], or to a long photocarrier lifetime $\tau$ due to the trapping of photoexcited carriers either in the defect states (persistent photoconductivity[28]) or in the band-bending region between a metal contact and a semiconductor[49]. However, as mentioned above, clean suspended MoS$_2$ devices only start to conduct ($G \sim 10^{-7}$ S) at large ($V_{ds}>E_g/e$) source-drain bias (Fig. 1b). Operation in this regime may be complicated by additional effects, such as Zener or thermal breakdown[49]. On the other hand, we observe that glass-supported MoS$_2$ devices (chosen to eliminate parasitic photogating) have dark conductance $G\sim 10^{-5}$ S, likely due to the higher doping level of supported MoS$_2$. In agreement with Eq. (1), the photoresponse of these devices is correspondingly higher and can be observed even at small $V_{ds}$ (Fig. 4b). Moreover, the relatively low resistance and correspondingly low RC time-constant of glass-supported devices allows us to measure the time dependence of the photocurrent.

The observation of $\eta \sim 25$ at $V_{ds}\sim 0.5$ V for a glass-supported device (Fig. 4b) rules out the avalanche effect as the mechanism responsible for the observed high photogain. In this regime, the energy $eV_{ds}$ is well below the fundamental band gap and is not sufficient to start an avalanche. Persistent photoconductivity has been previously reported in MoS$_2$[17], but we can exclude it as a possible candidate for the PC generation in clean MoS$_2$ because we routinely observe characteristic photoresponse time <1 ms at low temperatures (Fig. 4c). This is approximately five orders of magnitude faster than the response time reported for persistent photoconductivity[17]. The large photogain of our devices is most consistent trapping of one type of photocarriers due to band bending near the contacts[49]. Spatial separation of photocarriers (Fig. 4d) precludes their recombination, leading to long $\tau$, and correspondingly high photogain.

In conclusion, we note several potential applications of the obtained results. First, the large photogain, fast photoresponse, and bias-voltage dependence of the photocurrent spectra of pristine monolayer TMDCs suggest applications of these materials as sensitive and voltage-tunable photodetectors[50]. Second, the high absorption and dissociation probability of C-excitons may be employed in creating efficient TMDC-based solar cells[18,51]. Finally, suppression of disorder in monolayer TMDCs poses the question of intrinsic mobility of these materials[19].

**References:**

1. Castro Neto, A. H., Guinea, F., Peres, N. M. R., Novoselov, K. S. & Geim, A. K. The electronic properties of graphene. *Reviews of Modern Physics* **81**, 109-162 (2009).
2. Mak, K. F., Lee, C., Hone, J., Shan, J. & Heinz, T. F. Atomically Thin $MoS_2$: A New Direct-Gap Semiconductor. *Phys Rev Lett* **105**, 136805 (2010).
3. Xiao, D., Liu, G. B., Feng, W. X., Xu, X. D. & Yao, W. Coupled Spin and Valley Physics in Monolayers of $MoS_2$ and Other Group-VI Dichalcogenides. *Phys Rev Lett* **108** 196802 (2012).
4. Splendiani, A. *et al.* Emerging Photoluminescence in Monolayer $MoS_2$. *Nano Lett* **10**, 1271-1275 (2010).
5. Zeng, H. L., Dai, J. F., Yao, W., Xiao, D. & Cui, X. D. Valley polarization in $MoS_2$ monolayers by optical pumping. *Nat Nanotechnol* **7**, 490-493 (2012).
6. Mak, K. F., He, K. L., Shan, J. & Heinz, T. F. Control of valley polarization in monolayer $MoS_2$ by optical helicity. *Nat Nanotechnol* **7**, 494-498 (2012).
7. Sallen, G. *et al.* Robust optical emission polarization in $MoS_2$ monolayers through selective valley excitation. *Physical Review B* **86**, 081301 (2012).
8. Cao, T. *et al.* Valley-selective circular dichroism of monolayer molybdenum disulphide. *Nat Commun* **3**, 887 (2012).
9. Mak, K. F. *et al.* Tightly bound trions in monolayer $MoS_2$. *Nat Mater* **12**, 207-211 (2013).
10. Ross, J. S. *et al.* Electrical control of neutral and charged excitons in a monolayer semiconductor. *Nat Commun* **4** 1474 (2013).
11. Newaz, A. K. M. *et al.* Electrical control of optical properties of monolayer $MoS_2$. *Solid State Communications* **155**, 49-52 (2013).
12. Li, X., Zhang, F. & Niu, Q. Unconventional Quantum Hall Effect and Tunable Spin Hall Effect in Dirac Materials: Application to an Isolated $MoS_2$ Trilayer. *Phys Rev Lett* **110**, 066803 (2013).
13. Roldán, R., Cappelluti, E. & Guinea, F. Interactions and superconductivity in heavily doped $MoS_2$. *Physical Review B* **88**, 054515 (2013).
14. Geim, A. K. & Grigorieva, I. V. Van der Waals heterostructures. *Nature* **499**, 419-425 (2013).
15. Wigner, E. On the Interaction of Electrons in Metals. *Physical Review* **46**, 1002-1011 (1934).
16. Britnell, L. *et al.* Strong Light-Matter Interactions in Heterostructures of Atomically Thin Films. *Science* **340**, 1311-1314 10.1126/science.1235547 (2013).
17. Lopez-Sanchez, O., Lembke, D., Kayci, M., Radenovic, A. & Kis, A. Ultrasensitive photodetectors based on monolayer $MoS_2$. *Nat Nanotechnol* **8**, 497-501 (2013).
18. Bernardi, M., Palummo, M. & Grossman, J. C. Extraordinary Sunlight Absorption and One Nanometer Thick Photovoltaics Using Two-Dimensional Monolayer Materials. *Nano Lett* **13**, 3664-3670 (2013).
19. Wang, Q. H., Kalantar-Zadeh, K., Kis, A., Coleman, J. N. & Strano, M. S. Electronics and optoelectronics of two-dimensional transition metal dichalcogenides. *Nat Nano* **7**, 699-712 (2012).
20. Bolotin, K. I. *et al.* Ultrahigh electron mobility in suspended graphene. *Solid State Communications* **146**, 351-355 (2008).
21. Jin, T., Kang, J., Su Kim, E., Lee, S. & Lee, C. Suspended single-layer MoS2 devices. *Journal of Applied Physics* **114**, 164509 (2013).

**Acknowledgements:**
We thank Jed Ziegler and Richard Haglund for their help with optical measurements and acknowledge stimulating discussions with Tony Heinz. K.I.B. acknowledges support from ONR- N000141310299, NSF CAREER DMR-1056859, HDTRA1-10-0047, and Vanderbilt University. N.H.T would like to acknowledge support from DOE/BES and ARO through grant numbers FGO2-99ER45781 and W911NF-07-R-0003-02. Samples for this work were prepared at the Vanderbilt Institute of Nanoscale Science and Engineering using facilities renovated under NSF ARI-R2 DMR-0963361 and NSF EPS1004083. N.J.G., J.Y., D.M., and S.T.P. were supported by US DoE, BES, Materials Sciences and Engineering Division.

**Author Contribution:**
A.R.K. and A.K.M.N. prepared the samples, performed the experiment and analyzed the data; K.I.B supervised the project; B.W. and S.T.P. conducted the first-principles calculations; K.A.V. performed analytical calculations; D. P. prepared the glass-supported TMDC samples; A.R.K, A.K.M.N., H.K., B.L.I. and N.H.T. designed and built the spectroscopic measurement unit; D.C. A.B., N.J.G., J.Y. and D.G.M. grew the TMDC bulk crystals; A.R.K., A.K.M.N. and K.I.B. co-wrote the manuscript with input from all authors. All authors discussed the results.

# Supplementary Information

# Probing excitonic states in ultraclean suspended two-dimensional semiconductors by photocurrent spectroscopy

A. R. Klots[1]*, A. K. M. Newaz[1]*, Bin Wang[1], D. Prasai[2], H. Krzyzanowska[1], D. Caudel[1], N. J. Ghimire[3,4], J. Yan[4,5], B. L. Ivanov[1], K. A. Velizhanin[6], A. Burger[7], D. G. Mandrus[3,4,5], N. H. Tolk[1], S. T. Pantelides[1,4], and K. I. Bolotin[1]

[1]Department of Physics and Astronomy, Vanderbilt University, Nashville, TN-37235, USA
[2]Interdisciplinary Graduate Program in Materials Science, Vanderbilt University, Nashville, TN-37234, USA
[3]Department of Physics and Astronomy, University of Tennessee, Knoxville, TN-37996, USA
[4]Materials Science and Technology Division, Oak Ridge National Laboratory, Oak Ridge, TN-37831, USA
[5]Department of Materials Science and Engineering, University of Tennessee, Knoxville, TN-37996, USA
[6]Theoretical Division, Los Alamos National Laboratory, Los Alamos, NM-87545, USA
[7] Department of Physics, Fisk University, Nashville, TN-37208, USA

(*These authors contributed equally to the work)

S1. Fabrication and measurement techniques

S2. Intrinsic and extrinsic features in PC spectra

S3. Comparison of absorption and photocurrent spectra

S4. Temperature dependence of the excitonic peak positions

S5. Computational methods

S6. 2D Hydrogen model for band edge excitons

## S1. Fabrication and measurement techniques

*Sample Preparation*

Monolayer and multilayer TMDC samples were mechanically exfoliated from bulk TMDC crystals onto heavily *p*-doped silicon substrates covered by a 280 nm of thermal oxide. The sample thickness was confirmed by Raman spectroscopy measurements[1]. The metal electrodes were patterned by using e-beam lithography followed by thermal evaporation of Cr (2nm)/Au (90nm). The samples on glass were prepared by first fabricating Cr/Au electrode on top of a clean glass substrate and then by mechanically transferring TMDC flakes onto the electrodes[2]. To create suspended devices, the sacrificial $SiO_2$ was removed by buffered oxide etchant. To avoid collapse of suspended devices, finished specimens were dried in a critical point dryer[3].

*Optical measurements*

The samples were measured in an optical cryostat at temperatures between 70K and 300K. For photocurrent spectroscopy, the optical beam from a thermal source (halogen lamp) was guided through a monochromator (Cornerstone, Newport Corp.) and a mechanical chopper onto the sample where it was focused down to ~2 mm$^2$ spot. To calibrate the light intensity at a sample, the intensity of the split beam was recorded by a Si detector (Thorlabs DET36A). Both photocurrent and light intensity were measured simultaneously using a lock-in technique.

*Annealing of suspended devices*

Suspended monolayer devices are loaded into an optical cryostat with a base temperature ~77K. The devices are heated by an infrared (wavelength 10.6 μm) laser beam, while the cryostat is kept at its base temperature. To avoid structural and chemical modification of the sample, the laser beam is defocused into a ~0.5 mm$^2$ spot, so that the power flux on the sample is relatively low (<20μW/μm$^2$). We estimate that this procedure allows heating the die containing suspended devices by ≤500K.

During the process of annealing, the device is heated gradually by slowly increasing the laser power, and then measured using PC spectroscopy. The process is repeated until the background photocurrent below the A peak is no longer observed. We note that rarely (<5% of all measured devices), we observe signatures of low disorder – zero background PC below the A peak and between B and C peaks in **unannealed** devices. Spectra and electro-optical characteristics of one such device are presented in Fig. 2a and Fig. 4a-b. Similarity between PC spectra of clean unannealed (Fig. 2a) and clean annealed (Fig. 3a) devices proves that laser annealing does not chemically modify or damage the suspended monolayer materials.

For multilayer $MoS_2$ devices, suspension and annealing do not significantly reduce broad PC background. Because of this, the data reported in the main text (Fig. 3a) are collected using a device supported on a glass substrate. Due to their low surface-to-volume ratio, thick multilayer (>50L) flakes are not strongly affected by the environment. Therefore, we believe that the optoelectrical properties of such devices are close to intrinsic.

*Supported devices on glass*

To prepare samples on glass substrates, we first pattern Cr/Au electrodes on top of a microscopy glass slide (Corning Boro-Aluminosilicate, Delta-Technologies Limited). We start the preparation of the glass slides by cleaning with piranha solution (67% $H_2SO_4$ and 33% $H_2O_2$). After deposition of the electrodes, the substrate is again cleaned in piranha solution followed by reactive ion etching. Flakes of TMDC materials are then mechanically transferred onto electrodes[2].

*Lock-in measurements*

In our photocurrent measurements, monochromated light from a thermal source chopped at 930 Hz is used for illumination. The constant source-drain bias $V_{ds}$ up to 20 V is applied across the device. The absolute magnitude of photocurrent is recorded by measuring the source drain current with a lock-in amplifier. To remove the features due to artifacts, photocurrent at $V_{ds}=0$ is subtracted from the measured data (see below, section S2). Finally, the photocurrent is normalized to the illumination intensity (see below).

*Normalization of the photocurrent spectra*

To account for the variation of the spectral power density of the tungsten-halogen light source, we normalize the recorded PC spectra to the illumination power. For monolayer and bulk TMDC devices, the PC response is proportional to the illumination power and the procedure is justified (Fig. S1). For bilayer $MoS_2$ devices, however, the dependence between PC and illumination power is non-linear for the illumination intensity that we use. Because of this, we did not normalize the PC spectrum for a bilayer $MoS_2$ device that is shown in Fig. 3a. Non-uniform spectral density of the illumination source can significantly distort relative heights of A-, B- and C- peaks, but not their positions. Indeed, measured positions of the A and B peaks for bilayer $MoS_2$ are in close agreement with the positions of the peaks in optical absorption reported by Mak *et al.*[4].

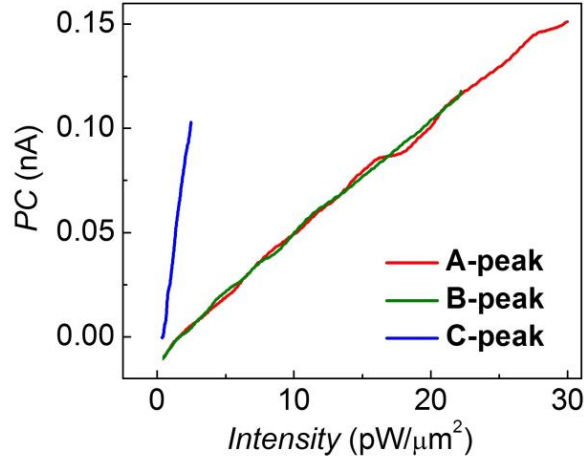

*Figure S1. Power dependence of the photocurrent in 1L $MoS_2$ device. Measurements were performed for the three observed peaks: A-peak at ~1.9 eV, B-peak at ~2.1 eV and C- peak at ~2.9 eV. We note that the photosensitivity at the C-peak wavelength is an order of magnitude larger than that for A- and B- peaks.*

*Device Statistics*

Table 1 presents the overall statistics of measured devices. In this table, the number outside the parenthesis is the number of successful devices, defined as the devices for which the PC spectra exhibits clearly resolved A-, B- and C-peaks with zero background below the A-peak. The number inside the parenthesis is the total number of fabricated devices.

*Table 1: Device Statistics*

|  | 1L $MoS_2$ | 2L $MoS_2$ | Multilayer $MoS_2$ | 1L $MoSe_2$ | 1L $WSe_2$ |
|---|---|---|---|---|---|
| Suspended on $Si/SiO_2$ wafer | 6(44) | 5(37) | 0(16) | 2(16) | 1(20) |
| Glass supported | 6(6) | 1(1) | 5(5) | - | - |

## S2. Intrinsic and extrinsic features in PC spectra

Photocurrent spectra of some devices show features that vary from device to device and hence cannot be interpreted as intrinsic spectral features of TMDCs. In this section, we analyze various mechanisms that produce artifacts in PC measurements. First, in supported and suspended devices before annealing (Fig. 1c, Fig. S2a), we observe broad featureless photocurrent spectrum with two dips (at ~1.9 eV and ~2.1 eV). The amplitude of the background photocurrent grows monotonically with $V_{ds}$. We attribute this background PC response to photo-field-effect (PFE)[5]. Second, in devices on Si/SiO$_2$, we observe non-zero photocurrent at zero $V_{ds}$ when lock-in measurements are used (Fig. S2a,b). This photocurrent signal at $V_{ds}$=0 is also largely featureless. We attribute this signal to displacement photocurrent[6]. Finally, sometimes we observe photocurrent between A and B peaks and above the B peak in devices where PFE and displacement photocurrent are not present (Fig. S2c). This photoresponse is likely due to **absorption by midgap states**[7-9] of TMDCs. Next, we will discuss the details of each of these mechanisms and outline the procedures for removing them.

*Photo-field-effect*

Current across a device under illumination is given by:

$$I = V_{ds} G(V_{ds}, n) \quad (S1)$$

, where $G$ is the device conductance, and $n$ is the number of carriers (see derivation of the Eq. (1) of the main text). In interpreting the PC spectroscopy data, we typically assume that the photocurrent originates from light-induced creation of additional photoexcited carriers inside a device. However, there is another less obvious way through which light can also produce changes in $I$. This mechanism, typically called photo-field-effect (PFE) or photogating effect[5], is effective when a light-absorbing semiconductor, such as Si substrate, is present close to a device. When the device is illuminated, photocarriers inside Si produce surface photovoltage at the Si/SiO$_2$ interface[10]. The resulting electric field effectively gates devices causing changes in $n$ and hence photocurrent. The following evidence indicates the contribution of photo-field-effect in our experiments:

1) As expected for PFE, the photocurrent response in TMDC devices on Si/SiO$_2$ features a broad background starting at ~1eV, which is the absorption edge of Silicon (Fig. S2a).
2) The amplitude of this background PC grows monotonically with $V_{ds}$, as expected for PFE (Eq. (S1)).
3) The background PC below the A-exciton vanishes completely for devices without light-absorbing substrate, such as MoS$_2$ devices on glass (Fig. S2c).

We note that photo-field-effect also disappears for suspended devices upon annealing (Fig. S2b). We do not currently understand the mechanisms of this effect.

Combined contribution of intrinsic photoresponse of MoS$_2$ and PFE can explain why A- and B- excitonic states in supported and suspended but unannealed devices appear as dips rather than peaks in PC measurements (Fig. 1c). Indeed, the sign of the surface photovoltage – and hence of PFE – is determined by the type of majority carriers in Si and can be opposite from the sign of the intrinsic photoresponse of MoS$_2$. Our lock-in measurements used in recording PC are only sensitive to the absolute value of the total photocurrent. If PFE signal is stronger than the intrinsic MoS$_2$ photoresponse, resulting measured current is $|I_{result}|=|I_{intrinsic}-I_{PFE}|= |I_{PFE}|- |I_{intrinsic}|$. In this situation, we expect to see excitonic photocurrent to appear as dips, rather than peaks, in agreement with the experimental measurements as shown in Fig.1c. We note that similar dips in photocurrent at the spectral positions of the A- and B- excitons have been previously observed in multilayer TMDCs by Kam *et al.*[11] However, the mechanism we propose here is different from that of Kam *et al.*

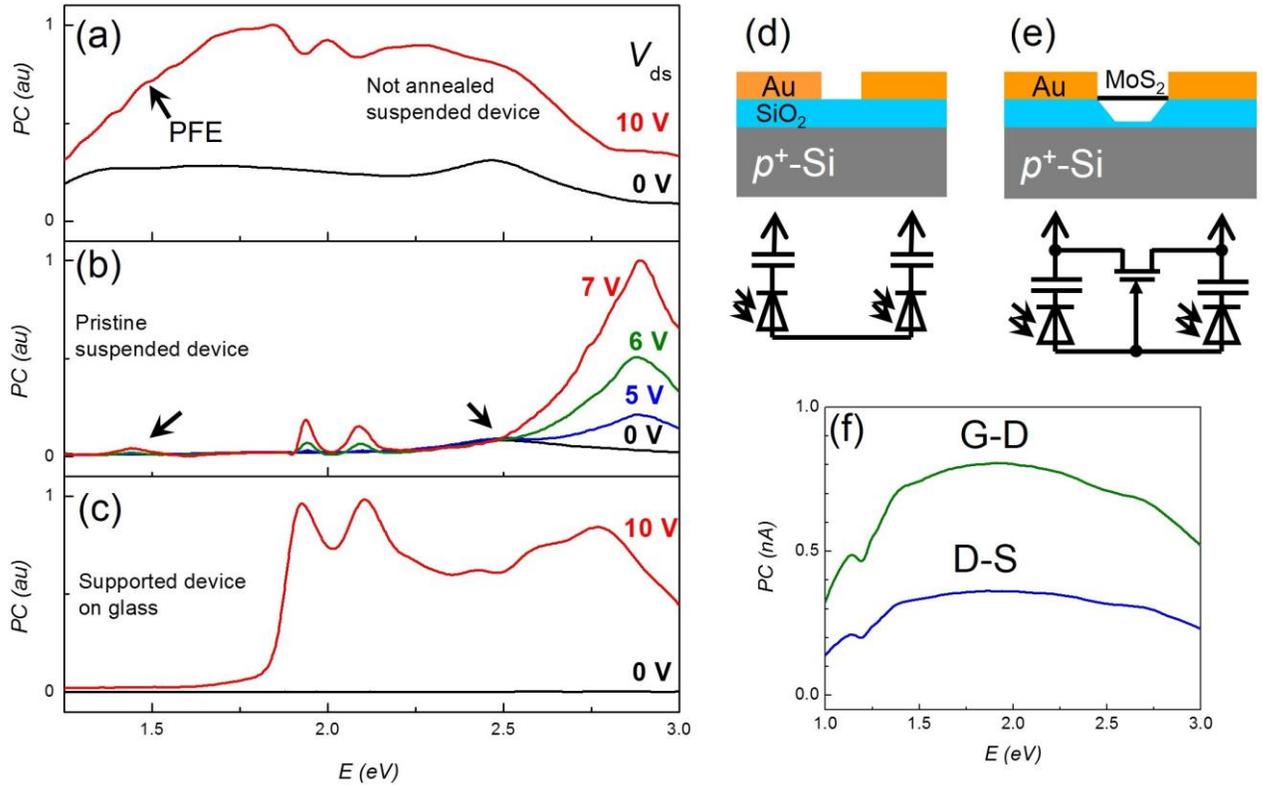

Figure S2: *Artifacts in PC of MoS$_2$-devices. (a)-(c) Photocurrent spectra for MoS$_2$ devices in different conditions. (a) Strong featureless background PC in suspended unannealed device due to photogating. (b) Pristine suspended unannealed device shows response at zero source-drain bias (black curve). (c) Device on glass shows no substrate-related artifacts, but has response in the region 2.0 eV-2.5 eV, likely due to the absorption by midgap states. (d) Structure and schematic representation of an empty device. Photodiodes symbolize photogeneration of surface charge on Si/SiO$_2$-interface and capacitors indicate the capacitance between Au electrodes and Si-substrate. (e) Structure of MoS$_2$-photodetector on Si-Substrate and its schematic representation. MoS$_2$ is represented as a field-effect-transistor, gated by the surface photovoltage on Si/SiO$_2$ interface. (f) Photoresponse of an empty device. Green curve: PC measured between gate and drain (G-D). Blue curve: PC measured between source and drain (D-S).*

*Displacement photocurrent*

For some devices on Si/SiO$_2$ wafers, we observe photocurrent even at zero source-drain bias (Fig. S2a-b). Moreover, similar PC response is observed for a test device that does not contain a conductive TMDC layer between the electrodes (Fig. S2f). Clearly, this type of photoresponse cannot be attributed to photo-field-effect and is not an intrinsic response of TMDCs. We propose that this photoresponse originates from the displacement photocurrent between Si and metal electrodes[6]. Indeed, our lock-in measurements of the photocurrent employ a light source with time-varying intensity. This produces a time dependent surface photovoltage of Si, and therefore time dependent potential difference across capacitor formed between Si back-gate and electrodes. Slight asymmetry between source and drain electrodes leads to the appearance of the displacement current between the electrodes. The following experimental evidence is consistent with the contribution due to displacement photocurrent:

1) Similarly to the case of photo-field-effect, displacement photocurrent vanishes for $MoS_2$ devices supported on glass (Fig. S2c, black curve). This is expected, as these devices do not contain silicon substrate.
2) As expected for displacement current, the phase of the PC at zero bias is shifted by ~90° with respect to the illumination signal.

Whereas photogating can be remedied by annealing a suspended device, displacement photocurrent is independent of the TMDC quality and cannot be completely eliminated by annealing. However, it can be removed by subtracting zero-$V_{ds}$ PC spectrum from the spectrum obtained under non-zero $V_{ds}$ bias.

*Disorder-related photocurrent*

As discussed above, both photo-field-effect and displacement photocurrent are fully suppressed in $MoS_2$ devices on glass. However, even in these devices, we observe the background PC response above the A peak that varies from device to device (Fig. S2c). At the same time, this background almost completely disappears in some suspended and annealed devices (Fig. 2a). This suggests that this background signal is *not intrinsic* to $MoS_2$. We propose that it is related to the absorption of light by the midgap states in TMDCs. Indeed, both experimental[7-9] and theoretical[12] work showed that disorder may modify optical spectra of TMDC and cause below-bandgap absorption.

To suppress disorder-related PC in a suspended device, we use thermal annealing. As discussed in the main text, background PC below the A peak and between B and C peaks disappears (Fig. 2a) upon annealing. This is consistent with reduction in disorder in thin TMDC films upon thermal annealing reported by Eda et al.[8]

## S3. Comparison of absorption and photocurrent spectra

Devices on glass, which are described above, also allow us to compare our photocurrent spectra with absorption spectra of $MoS_2$ obtained by other groups[4]. The photocurrent spectrum of a glass-supported 1L $MoS_2$ device, as shown in Fig. S3, is very similar to the optical absorption spectrum of 1L $MoS_2$ devices measured under similar conditions by Mak *et.al.*[4] This suggests that the photocurrent spectrum is roughly proportional to optical absorption spectrum and probes similar transitions. This is consistent with our expectation that the photogain (See Eq.1 of the main text) does not strongly depend on wavelength.

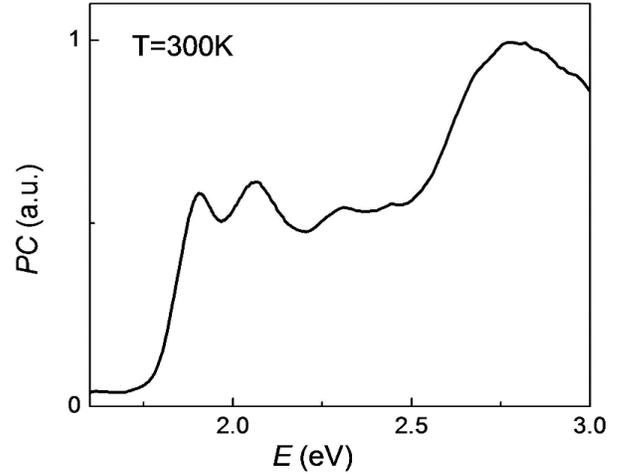

*Figure S3: PC spectrum of a glass-supported 1L $MoS_2$ device at room temperature.*

## S4. Temperature dependence of the excitonic peak positions

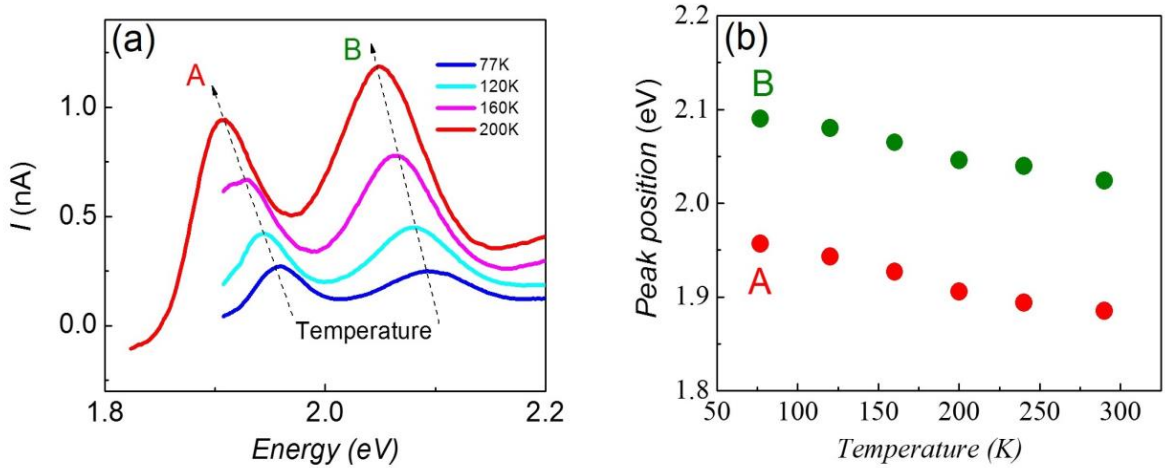

*Figure S4: Temperature dependence of A- and B- peaks. (a) Spectra of a suspended 1L MoS$_2$ device at different temperature. (b) A- and B- peak positions of the same device as a function of temperature.*

In our experiments, the positions of the A- and B- excitonic peaks in a suspended device blue-shift by ~150meV as we cool down the device from 300K to 77K (Fig. S4). Similar increase of MoS$_2$ bandgap with decreasing temperature has been previously reported[13]. In order to compare our PC spectrum recorded at 77K to the PL spectrum recorded at room temperature, we blue-shift the PL-spectrum by 150meV as in Fig. 1e.

## S5. Computational methods

Density functional calculations are performed using the VASP package[14]. The PBE-GGA exchange-correlation potential[15] is used and electron-core interactions are treated in the projector augmented wave (PAW) method[16,17]. The plane-wave kinetic-energy cutoff is set to 400 eV. The spin-orbit coupling is taken into account. The layered MoS$_2$ is modeled using a supercell approach with adjacent layers separated by a vacuum region. The k-point mesh is set to $12 \times 12 \times 1$ or higher (see the discussion below). The structures are optimized with all the atoms relaxed until the self-consistent forces reached 0.02 eVÅ$^{-1}$. The van der Waals interactions are taken into account using the semi-empirical correction scheme, the DFT+D approach[18].

The GW$_0$ calculations are performed using VASP[19]. The quasiparticle energies are obtained by iterating only Green function (G), but keeping screened Coulomb interactions (W) fixed to the initial DFT W$_0$. The energy cutoff for the response function is set to 100 eV to speed up the calculations, and about 100 empty bands are included in the calculations. The obtained quasiparticle energies are then employed to calculate the band structure using Wannier interpolation as implemented in WANNIER90 program[20]. Binding energies of the direct excitons are calculated by solving the Bethe-Salpeter equation (BSE) for the two-particles Green's function[21,22]. The calculations of BSE spectrum are carried out based on the Tamm-Dancoff approximation[23]. Eight highest valence bands and the eight lowest conduction bands are included in the calculation of the spectrum.

Multiple factors impact the precision of GW-BSE calculations. It has been shown[24] that for modeling mono- and few-layer TMDC materials, most relevant parameters are (i) k-point sampling of the Brillouin zone (BZ), (ii) vacuum layer thickness – spatial separation between two next neighboring TMDC layers (effectively, the height of a supercell), and (iii) number of GW-iterations. Unfortunately, limitations of the computational resources make it impractical to achieve convergence of the spectrum with respect to all computational parameters.

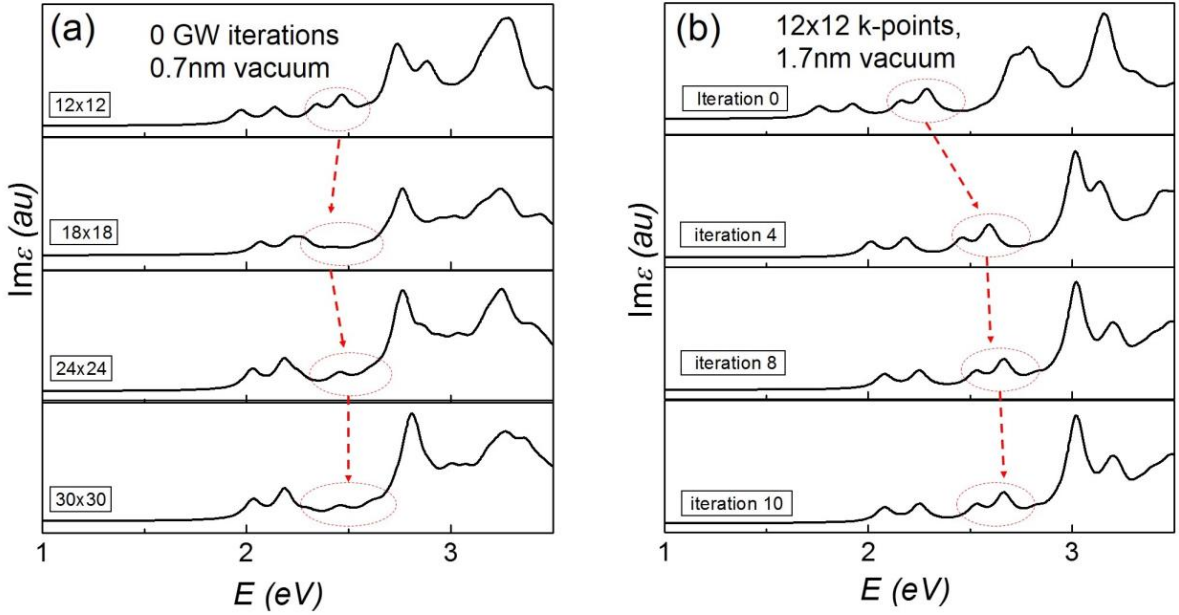

*Figure S5: First-principles calculations for MoS$_2$ absorption spectra with different parameters. (a) Models of MoS$_2$ optical spectra with different k-point grids, no GW-iterations and small vacuum spacing. (b) Evolution of MoS$_2$ optical spectrum with iterations of the Green's function (G). These models are based on small k-point grid and large vacuum spacing. Red circles in (a) and (b) together with red arrows highlight the evolution of artificial features with k-point sampling/GW-iterations.*

In order to choose optimal modeling parameters, we now discuss qualitatively how the factors discussed above affect the computational results.

i. *k-point sampling*: The k-point sampling $n \times n$ of a BZ allows us to map the particle's wavefunction to a real space area of only $n \times n$ unit cells. Thus, BZ with small sampling may cause artificial confinement of the exciton leading to artifacts in absorption spectrum (see below).

ii. *Vacuum layer thickness*: Using a periodic supercell in the calculation, modeling of 2-dimensional material is performed by its repetition in the *z*-direction with certain vacuum spacing. Too small vacuum layer results in increasing dielectric screening due to neighboring layers, which reduces the band gap and exciton binding energy.

iii. *Number of GW-iterations*: As GW-iterations are performed, we increase precision of particle's Green's function, which significantly affects particle's self-energy ($\Sigma = iGW$), the single-particle band structure and *e-h* interactions.

In order to perform accurate calculations within our computational resources, we have to optimize the most significant parameters. First, we calculate the optical spectrum using different k-points samplings as shown in (Fig. S5a). Since excitons in TMDCs are tightly bound, they are highly localized (within ~1 nm). A 12×12 grid corresponds to 12×12 unit cells in the real space, which corresponds to the confinement of excitons in lateral dimension of ~4 nm size. Thus, we expect that 12×12 k-point sampling will not significantly change the peak positions (Fig. S5a). We note that the features between ~2.3eV and ~2.7 eV (marked by red ovals) vary in amplitude and shape depending on the density of the mesh (Fig. S5a). We attribute these features to the artifacts of the calculation, which we discuss in details below.

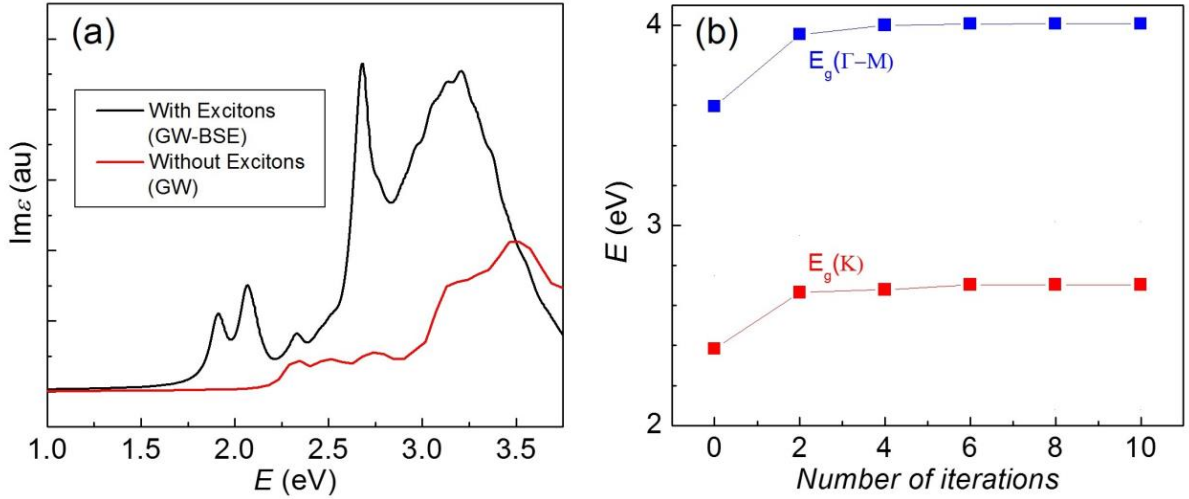

*Figure S6: Optical spectra of MoS$_2$ with and without excitonic effects and their convergence. (a) Optical spectra of MoS$_2$ with (black curve) and without (red curve) e-h interactions, performed for 24× 24 k-points, 0.7nm vacuum spacing and zero GW-iterations. We note that without e-h interactions, the absorption demonstrates a step-like increase at ~2.2eV (bandgap for this model is ~2.2 eV) as expected for hyperbolic bands at the K-valleys. Same features remain in place after inclusion of excitonic effects. (b) Evolution of the bandgap energies at different points of the BZ with GW$_0$ iterations. Red and blue squares indicate values of single-particle (GW$_0$) bandgap for K-point (location of A-exciton) and for the minimum between Γ- and M-points (one of the locations of C-exciton), respectively.*

When photon energy exceeds the band gap, BZ sampling-related spatial confinement will cause discreet peaks instead of continuous absorption[24]. As we increase k-point sampling, peaks are expected to merge into continuum as evident from Fig. S5a. This is seen clearer in Fig. S6a, where we compare optical spectra calculated with and without excitonic effects using a (24×24) k-point sampling in a G$_0$W$_0$ calculation. The calculation without excitonic effects exhibits a plateau, starting at the band gap value ~2.2eV, and a residual peak at ~2.3 eV. These features remain in place after inclusion of the excitonic effects (via BSE), *i.e.*, these features are non-excitonic in nature. Hence, comparing spectra at different k-point sampling presented in Fig.S5a, we can distinguish between artificial ones (Fig. S5a, red dashed ovals) and real peaks.

Then, we increase vacuum layer to 17 Å and perform $GW_0$ calculations using a 12×12 k-point sampling (Fig. S5b) while tracking positions of real and artificial peaks. As evident from Fig. S5b and Fig. S6b, four iterations are necessary for the convergence of the band structure. We find that calculated A- and B-peak positions are higher than the experimental values by ~160 meV for $MoS_2$.

Therefore, A- and B-peak positions alone cannot serve as a good criteria to compare the GW-BSE spectrum with the experimental results. However, we can compare the relative positions of A-, B- and C- peaks ($E_C$-$E_A$, $E_C$-$E_B$) instead of their absolute positions ($E_A$, $E_B$ and $E_C$). The separation between these three peaks provides a basis for validating our calculations, *i.e.*, by comparing them with experimental data. To account for the mismatch between the measured and calculated peak positions (for example, ~160 meV for $MoS_2$), we manually red-shift the computational data to align calculated and experimental A- and B- peak positions (Fig. 2a, Fig. 3a).

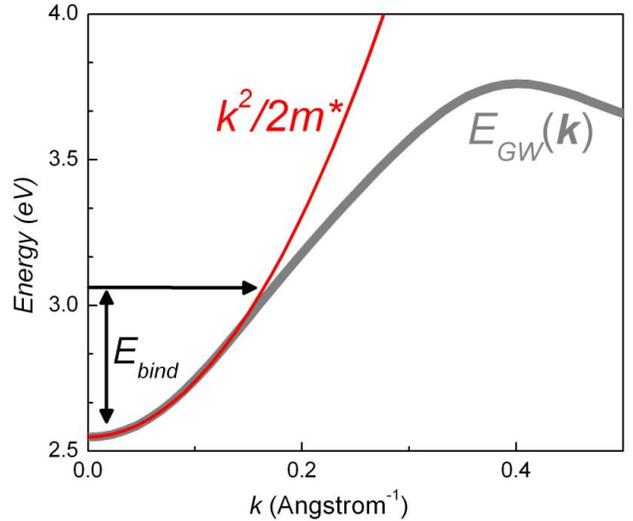

*Figure S7: 2D-hydrogen model for an exciton in TMDC. We compare first-principles calculated optical bandstructure with the parabolic dispersion, expected in a 2D-hydrogen approximation. Here m\* is a reduced mass of an e-h pair. Vertical black arrow indicates the magnitude of the exciton binding energy.*

**S6. 2D Hydrogen model for band edge excitons**

To estimate the binding energies and the oscillator strength of the ground and the excited states of excitons in TMDCs, we use a non-relativistic 2D hydrogen model. On the basis of our first-principle calculations, we estimate the reduced mass of an *e-h* pair $m^*$~$0.2m_e$ (here $m_e$ is the electron mass). If we assume the dielectric constant of $MoS_2$ to be ~4.2[25], the binding energy of an A-exciton, according to the 2D-hydrogen model with *1/r* potenital, can be estimated as[26]

$$E_{bind} = \frac{1}{2(n-1/2)^2} \frac{m^* e^4}{(4\pi\varepsilon)^2 \hbar^2},$$

giving the ground state binding energy $E_{bind}$~620 meV. This value agrees well with the experimentally estimated value ~570 meV. In the same model, the oscillator strength decreases[27] with *n* as $f_n \propto 1/(n-1/2)^3$. For example, the exciton's first excited state has oscillator strength smaller by a factor $f_1 / f_2 = 27$ than that for the ground state.

Finally, we discuss the validity of the 2D hydrogen approximation. In order to study the physics of excitons, it is convenient to work in the center-of-mass (CM) frame. In this frame of reference, the dispersion of particles constituting an exciton is given by the difference between the valence and the conduction bands[28] (*i.e.*, optical band structure). On the other hand, in a non-relativistic 2D-hydrogen model, the dispersion relation is parabolic. Thus, the non-relativistic 2D-hydrogen model is considered a good approximation if the calculated optical band structure is close to the parabolic dispersion at the exciton binding energies. Fig. S7 shows a calculated optical bandstructure (grey curve) of $MoS_2$ near the K- point, and its parabolic approximation based on the 2D-hydrogen model (red curve). At relevant energies ($E_g$+$E_{bind}$, marked by horizontal arrow), we see ~10% deviation of the parabolic approximation from the calculated GW dispersion. This implies that 2D-hydroden model is sufficient for qualitative treatment of the A/B-excitons.